\documentclass[aps,prl,reprint,twocolumn,showpacs,superscriptaddress,floatfix]{revtex4-1} 
\usepackage{graphicx}%
\usepackage{dcolumn}%
\usepackage{bm}%
\usepackage{latexsym}
\usepackage{amsmath}
\usepackage{natbib}

\def\be{\begin{equation}}
\def\ee{\end{equation}}
\def\bea{\begin{eqnarray}}
\def\eea{\end{eqnarray}}

\renewcommand{\Re}{\,\mathrm{Re}}
\renewcommand{\Im}{\,\mathrm{Im}}

\begin{document}

\title{The valence electron photoemission spectrum of semiconductors: ab initio description of multiple satellites}

\author{Matteo Guzzo}
\email[]{matteo.guzzo@polytechnique.edu}

\author{Giovanna Lani}

\author{Francesco Sottile} 
\affiliation{Laboratoire des Solides Irradi\'es, \'Ecole Polytechnique, CNRS, CEA-DSM, F-91128 Palaiseau, France}
\affiliation{European Theoretical Spectroscopy Facility (ETSF)}

\author{Pina Romaniello} 
\affiliation{Laboratoire de Physique Th\'eorique, CNRS, Universit\'e Paul Sabatier, F-31062 Toulouse, France}
\affiliation{European Theoretical Spectroscopy Facility (ETSF)}

\author{Matteo Gatti} 
\affiliation{ Departamento F\'isica de Materiales, CSIC-UPV/EHU-MPC and DIPC,
Universidad del Pa\'is Vasco, E-20018 San Sebasti\'an, Spain}
\affiliation{European Theoretical Spectroscopy Facility (ETSF)}

\author{Joshua J. Kas} 
\altaffiliation{}
\affiliation{Department of Physics, University of Washington, Seattle, WA 98195}

\author{John J. Rehr} 

\affiliation{Department of Physics, University of Washington, Seattle, WA 98195}
\affiliation{European Theoretical Spectroscopy Facility (ETSF)}

\author{Mathieu G. Silly} 

\author{Fausto Sirotti} 
\affiliation{Synchrotron-SOLEIL, BP 48, Saint-Aubin, F91192 Gif sur Yvette CEDEX, France}

\author{Lucia Reining}
\email[]{lucia.reining@polytechnique.fr}

\affiliation{Laboratoire des Solides Irradi\'es, \'Ecole Polytechnique, CNRS, CEA-DSM, F-91128 Palaiseau, France}
\affiliation{European Theoretical Spectroscopy Facility (ETSF)}

\date{\today}

\begin{abstract}
The experimental valence band photoemission spectrum of semiconductors exhibits multiple satellites 
that cannot be described by the GW 
approximation for the self-energy in the framework of many-body perturbation theory. 
Taking silicon as a prototypical example, we compare experimental high energy photoemission spectra 
with GW calculations and analyze the origin of the GW failure. 
We then propose an approximation to the functional differential equation that determines the exact 
one-body Green's function, whose solution has an exponential form. 
This yields a calculated spectrum, including cross sections, secondary electrons, and an estimate 
for extrinsic and interference effects, in excellent agreement with experiment. 
Our result can be recast as a dynamical vertex correction beyond GW, giving hints for
further developments. 
\end{abstract}

\maketitle

Photoemission is a prominent tool to access information about electronic structure 
and excitations in materials. Modern synchrotron sources can provide detailed insight, 
thanks to their high intensity and broad photon energy range. But the interpretation of the 
experimental data is far from obvious, and theory is an essential complementary tool. 
However, \emph{ab initio} calculations typically focus on bulk bandstructure 
\cite{rep.prog.phys.61.237,ssp.54.1}; thus surface effects
are ignored, and satellites are not included. 
The latter are a pure many-body effect
due to coupling to excitations of the material.
Such many-body effects are contained in approaches developed for correlated
materials \cite{georges96,kotliar06} however, these are usually based on models with short-range 
interactions, whereas satellites such as plasmons involve long-range effects.  
Plasmon satellites have been extensively studied in core-level experiments 
\cite{[{e.g.\@: }][{}]PhysRevB.76.085422}. 
There they can be described by a theoretical model where a single dispersionless 
fermion couples to bosons. 
The resulting exact Green's function
has an exponential form given by the so-called cumulant expansion (CE). 
A Taylor expansion of the exponential leads to a well defined quasi-particle (QP) peak followed 
by a decaying series of plasmon satellites at energy differences given by the
plasmon energy, consistent with experimental observations 
\cite{HEDIN80,Hedin99,NOZIERES69,LANGRETH70,Bechstedt2003}.  
In the valence region, plasmon satellites are much less studied, though  
\emph{ab initio} approaches can provide a good starting point.
At high photoelectron energies
the photoemission spectrum is approximately proportional to the intrinsic spectral function
$A(\omega) = -(1/\pi){\rm Im}\, \mathcal{G}(\omega)$, where $\mathcal{G}$ is
the one-particle Green's function. The latter is typically calculated using the widely used GW 
approximation (GWA) \cite{Hedin65,Hedin69,Hedin99}. 
In principle, the GWA contains correlations effects beyond the
quasiparticle approximation.  
However, these additional features are rarely calculated due to computational complexity and, more 
importantly, the serious discrepancies between GWA 
and experiment (see e.g.\@ \cite{PhysRevB.46.13051,Aryasetiawan96,ISI:000179080800133,Kheifets03}).
The CE has also been used for the homogeneous electron gas \cite{Holm97}
and simple metals \cite{ISI:000179080800133,Aryasetiawan96}, yielding an improved description
of satellites over GW.  
Silicon \cite{Kheifets03} and graphite \cite{ISI:000166608600008} were also studied, but no plasmon 
satellite series were observed. However, these results are not conclusive due
to difficulties of interpreting the experimental data. 
This leaves a series of important questions: (i) do materials generally exhibit 
intrinsic satellites in the valence band region following a cumulant like distribution, or are the 
extrinsic plasmon peaks,
due to losses incurred by the escaping photoelectron, dominant?
(ii) if such series are seen, how bad are \emph{ab initio} GW 
calculations, what is the reason for their failure, and (iii) how can they be improved? 
Answering these questions would be a crucial step towards a better understanding of correlation 
effects in electronic excitations and a predictive \emph{ab initio} approach to 
photoemission.

In this work we focus on plasmon satellites
using silicon as a prototypical example.
We have obtained valence band photoemission data at high photon energy
(XPS) that constitute a reliable and well resolved benchmark. 
Analysis of the data allows us to elucidate the failure of GW in describing the 
satellites. 
Then, starting from the fundamental equations of many-body perturbation theory (MBPT), we show how 
the failure can be overcome by using a decoupling approximation that leads to an exponential 
representation of the one-particle Green's function. 
Together with an estimate for extrinsic and interference effects, we obtain
results for the quasiparticle peaks and satellites in excellent agreement
with experiment. Our theoretical results can be expressed in terms of a dynamical vertex correction, 
a powerful basis for further modelling.

Angular resolved valence photoemission (ARPES) measurements were performed at the UHV photoemission 
experimental station of the TEMPO beamline \cite{exp1} at the SOLEIL synchrotron radiation source.
Linearly polarized photons from the Apple II type Insertion Device (HU44) were selected in energy 
using a high resolution plane grating monochromator with a  
resolving power $E/\Delta E = 5000$.
The end-station chamber (base pressure $10^{-10}$ mbar) is equipped with a modified SCIENTA-200 
electron analyzer with a delay-line 2D detector which optimizes the detection linearity and 
signal/ background ratio \cite{ISI:000287530900019}.  
The  overall energy 
resolution was better than 200 meV.
The photon beam impinges on the sample at an angle of 43$^{\circ}$, and photoelectrons were 
detected around the sample surface normal with an angular acceptance of $\pm 6^{\circ}$. 
A $n$-type ($N_D\simeq 2\times 10^{-18} P$ atoms/cm$^3$) Si(001) wafer was cleaned from the native 
oxide by flash annealing at 1100$^\circ$ C after prolonged degassing at 600$^\circ$ C in ultra-high vacuum.  
The silicon surface was annealed at 300$^\circ$ C to prevent surface etching, and hydrogenated in a 
partial pressure of activated hydrogen about $2\times 10^{-8}$ mbar for 20 min.
The ARPES was measured along the $\Sigma$ direction.  
At 800 eV kinetic energy the Si Brillouin zone is observed with an emission angle slightly 
smaller than 5$^{\circ}$. 
The measured photoemission map was integrated over the spectral intensity originated by 
two Brillouin zones.  
The Fermi level was obtained by measuring a clean Au(111) surface.
\begin{figure}[!htbp]
   \includegraphics[width=1.0\columnwidth]{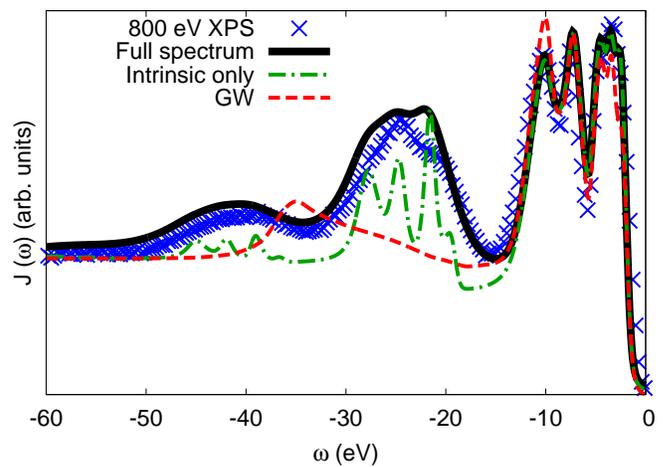}
\caption{(Color online) Experimental XPS  spectrum of Si at 800 eV photon energy (blue crosses), 
compared to the theoretical intrinsic $ A ( \omega ) $ 
calculated from $ G_0 W_0 $ (red dashed), and from Eq.\@ \eqref{eq:finalA} (green dot-dashed). 
On top of the latter the black solid line also includes extrinsic and interference effects.
All spectra contain photoabsorption cross sections, a calculated secondary electron background and 
0.4 eV Gaussian broadening to account for finite
$k$-point sampling and experimental resolution. 
The Fermi energy is set to 0 eV.}
\label{fig:spf-final}
\end{figure}
The experimental data (crosses) are summarized in 
Fig.\@ \ref{fig:spf-final}.
One can distinguish the quasiparticle peaks between the Fermi level at zero and the bottom valence 
at -12 eV, followed by two prominent satellite structures, each at a mutual distance
of about 17 eV, as well as a more weakly visible third satellite between -52 and -60 eV. 
These structures are obviously related to the 17 eV silicon bulk plasmon \cite{ISI:A1978FY99000003}. 

The exact one-electron Green's function $\mathcal G$ is described by an equation of motion
with the form of a functional differential equation \cite{kadanoffbaym},
\begin{equation}
 \mathcal{G}  = \mathcal{G}_{0} + \mathcal{G}_{0} V_H  
  \mathcal{G}  + \mathcal{G}_{0} \varphi \mathcal{G} 
  + i  \mathcal{G}_{0} v_c \frac {\delta \mathcal{G}  } {\delta \varphi } .
\label{eq:diffeq}
\end{equation}
Here $ \mathcal{G}_{0}$ is the non-interacting Green's function, $ \varphi $ is a fictitious 
external perturbation that is set to zero at the end of the 
derivation, $ v_c $ is the bare Coulomb interaction, and all quantities are understood to be 
matrices in space, spin, and time. The Hartree potential $V_H$ gives rise to screening to all 
orders. Linearizing $V_H$ with respect to $\varphi$ yields
\cite{2011arXiv1103.1630L}
\begin{multline}
 \mathcal{G}  (t_1 t_2) = \mathcal{G}_{H}^0 (t_1 t_2)  + \mathcal{G}_{H}^0 (t_1 t_3) 
  \bar \varphi  (t_3) \mathcal{G} (t_3 t_2) \\
  + i  \mathcal{G}_{H}^0 (t_1 t_3) \mathcal{W} (t_3 t_4) 
   \frac {\delta \mathcal{G}(t_3 t_2) } {\delta \bar \varphi (t_4) } , 
\label{eq:diffggqp}
\end{multline}
where $\bar \varphi $ is equal to $\varphi$ screened by the inverse dielectric function, 
$\mathcal{W}$ is the screened Coulomb interaction, and  $\mathcal{G}_{H}^0 $ is the Green's function 
containing the Hartree potential at vanishing $\bar \varphi $; only time arguments are displayed 
explicitly and repeated indices are integrated. 
This linearization preserves the main effects of $\mathcal{W}$ and hence of plasmons. 
With the additional approximation 
$ \frac {\delta \mathcal{G} (t_3 t_2) } { \delta \bar \varphi (t_4) } 
 \simeq \mathcal{G} (t_3 t_4)\mathcal{G} (t_4 t_2) $ one obtains the Dyson equation 
$ \mathcal{G} = \mathcal{G}_H^0 + \mathcal{G}_H^0 \Sigma \mathcal{G} $ in the GWA for the 
self-energy $\Sigma$. 
However this approximation can be problematic.
For the following analysis we use the standard $G_0W_0$ approach, where ${\mathcal G}_0$ is taken 
from an LDA calculation and ${\mathcal W}_0$ is the screened interaction in the Random Phase Approximation. 
%
\begin{figure}[!htbp]
   \includegraphics[width=1.0\columnwidth]{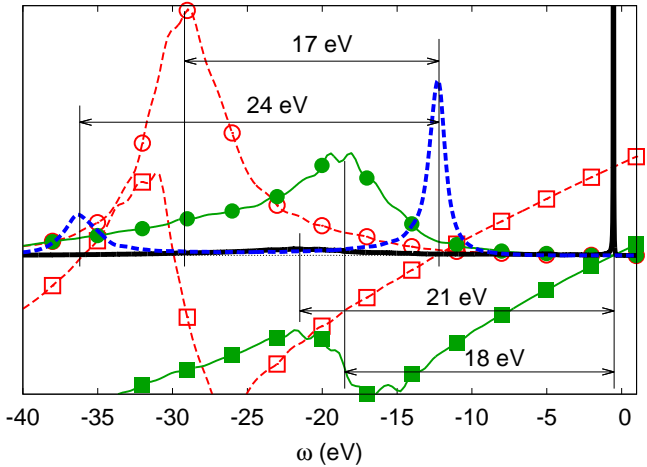}
\caption{(Color online) $G_0W_0$ spectral function of bulk silicon for the
top and bottom valence bands at the $\Gamma$ point
(black solid and blue dashed, respectively).
The corresponding imaginary parts of the self-energy 
(red empty-circles-dashed-line and green full-circles-solid-line) and $ \omega - \varepsilon_{ H } - \Re\Sigma $
 (red empty-squares-dashed-line and green full-squares-solid-line) are also shown.
The Fermi energy is set to 0 eV.}
\label{fig:spf-gw-b4}
\end{figure}
Fig.\@ \ref{fig:spf-gw-b4} 
shows the
$ G_0 W_0 $ spectral function \cite{Fleszar97}
$ A ( \omega ) =  \frac1\pi\; { \left| \mathrm{Im}\Sigma (\omega) \right| }/[
{ \left[ \omega - \varepsilon_H - \mathrm{Re}\Sigma (\omega) \right]^2 + 
\left[  \mathrm{Im}\Sigma (\omega) \right]^2 ]}
$
of Si at the $ \Gamma $ point, for top valence (solid line) and bottom valence (dashed), respectively. 
The top valence shows a sharp quasiparticle peak followed by a broad, weak satellite structure at about -21 eV.
This peak stems from the prominent peak in $ \Im\,\Sigma $ (full circles) at about
$ - 18 $ eV, itself due to the plasmon peak in $\Im \mathcal{W}$.
It is a typical plasmon satellite, 
though (cf.~\cite{Hedin99}), the QP-satellite spacing is slightly overestimated because the 
term $ \omega - \varepsilon_{ H } - \Re\Sigma $ (full squares) in the
denominator of the expression for $A(\omega)$ is not constant. 
However the GWA has a more severe problem: for the bottom 
valence, the satellite structure at about $ - 36 $ eV is much too far from the QP peak at about 
$ - 12 $ eV, and much too sharp. 
This satellite does not correspond to a plasmon peak in 
$ \Im\Sigma $ (empty circles), but to a zero in 
$ \omega - \varepsilon_{ H } - \Re\Sigma $ (empty squares)
in the denominator of $A(\omega)$, as for a QP peak.
It has been  interpreted in the HEG as a \emph{plasmaron}, 
a coupled hole-plasmon mode \cite{Lundqvist1967a,*ISI:A1968A818100003}, 
but as noted below it is an artifact of the GWA \cite{ISI:A1972N855800007,ISI:A1973P093700012}. 
Fig.\@ \ref{fig:spf-final} compares the total $ GW $ spectral function
(dashed red line) 
summed over all valence bands and $k$-points, with our XPS data. 
The effects of cross-sections are included by projecting on angular momenta
in atomic spheres using the atomic data of Ref.\@ \cite{Yeh85}.
The secondary electrons background at energy $\omega$ 
was determined by integrating the calculated intrinsic 
spectral intensity between $\omega$ and the Fermi level, 
similar to 
\cite{PhysRevB.5.4709}.
A constant scaling factor was set such that the 
measured photoemission intensity at the highest binding 
energy (60 eV), where primary electrons intensity is 
absent, is reproduced.
As expected, the dominant QP spectrum is well described by $GW$, but the
satellite is dominated by the plasmaron around $ - 36 $ eV, in complete
disagreement with experiment.  The experimental plasmon satellite at about
$-25$ eV  appears only as a weak shoulder in the $GWA$.
Thus the plasmaron peak is responsible for the GWA failure \cite{ISI:A1972N855800007,ISI:A1973P093700012} in silicon. 

To go beyond the GW self-energy requires vertex 
corrections. 
However, adiabatic vertex corrections (see e.g.\@ \cite{PhysRevB.49.8024}) 
only lead to renormalization of energies and do not 
create new structures. 
Thus alternatively, we concentrate here on dynamical 
effects, and we choose to approximate directly Eq.\@ \eqref{eq:diffggqp}, 
without passing through a self-energy.
We decouple Eq.\@ \eqref{eq:diffggqp} approximately
by supposing that $\mathcal{G} $ and $\mathcal{G}_H $ are diagonal in
the same single particle basis. 
Eq.\@ (\ref{eq:diffggqp}) is then applicable separately for every single
matrix element of $\mathcal{G}$ and each state couples independently to
the neutral excitations of the system through $\mathcal{W}$ 
\cite{[{In a similar spirit the equation of motion for the Green's function 
is decoupled in: }][{}]Almbladh83}. 
The latter can now be understood as the screened intra-orbital Coulomb matrix element for the 
chosen state. 
Such a decoupling approximation can be optimized \cite{ISI:A1972N855800007,ISI:A1973P093700012} by adding and subtracting a self-energy correction, 
hence by using a QP Green's function $\mathcal{G}_{\Delta}$ 
obtained from a good QP self-energy instead of $\mathcal{G}_H $. 
Since the GWA is currently the state-of-the art for QP properties, we suppose that for every 
decoupled state $k$, $ \mathcal{G}^k_{\Delta}( \tau ) = i \theta ( - \tau ) 
e^{ -i \varepsilon_k \tau } $ is determined from $ \Sigma^{GW}(\varepsilon_k)$,  where 
$ \varepsilon_k = \varepsilon^0_k + \Sigma^{GW} ( \varepsilon_k ) $ is the (complex) GW 
quasiparticle energy and $ \tau = t_1 - t_2 $. 
Now Eq.\@ \eqref{eq:diffggqp} can be solved exactly for each state. 
Briefly the main steps are: (i) solve the non-interacting ($\mathcal{W}=0$) version of 
\eqref{eq:diffggqp}, which leads to an explicit solution $\mathcal{G}_{\Delta}^{\varphi}$; 
(ii) iterate the result  $\mathcal{G}=\mathcal{G}_{\Delta}^{\varphi} - 
\mathcal{G}_{\Delta}^{\varphi} \Delta\mathcal{G} + i\mathcal{G}_{\Delta}^{\varphi} 
\mathcal{W} \frac {\delta \mathcal{G}  } {\delta \bar \varphi  } $ 
starting from $\mathcal{G}^{(0)}=\mathcal{G}_{\Delta}^{\varphi}$.  Here
$ \Delta$ compensates for the self-energy insertion used for the optimized decoupling;
(iii) use the exact relation
$ \frac {\delta \mathcal{G}_{\Delta}^{\varphi} (t_3 t_2) } {\delta \bar \varphi (t_4) } = 
\mathcal{G}_{\Delta}^{\varphi} (t_3 t_4) \mathcal{G}_{\Delta}^{\varphi} (t_4 t_2) = 
 i \mathcal{G}_{\Delta}^{\varphi} (t_3 t_2) \theta (t_2-t_4)\theta(t_4-t_1) $
to derive
\be
 \mathcal{G} ( t_1 t_2 ) = \mathcal{G}_\Delta ( \tau ) e^{ i \Delta\tau } 
  e^{ i \int_{t_1}^{t_2} dt' [ \bar \varphi(t') 
  - \int_{t'}^{t_2} dt'' \mathcal{W} ( t' t'' ) ] } .
\label{eq:onelevelg} 
\ee
The equilibrium solution is obtained setting
$\bar \varphi = 0$.

In silicon, where the peaks in the loss function are well defined, it is justified 
to use a single plasmon pole model 
$\mathcal{W} ( \tau ) = -i \lambda_k \left\{ e^{ - i \tilde{\omega}_k \tau } \theta ( \tau )
 + e^{ i\tilde{\omega}_k \tau } \theta ( - \tau ) \right\}$ 
 with plasmon energy $\tilde \omega_k$ and intrinsic strengths $\lambda_k$ for each matrix element of 
$ \mathcal{W}$. 
Besides $\bar \varphi$, the total exponent becomes then
$ a_k \left[ e^{ i \tilde\omega_k  \tau  } - 1 \right] $ 
with $a_k = \lambda_k/\tilde \omega_k^2$ obtained from the corresponding GW results. 
We find that $ a_k$ varies around 0.3.
Taylor expansion of the exponential leads then to the spectral function 
\be
 A_k( \omega ) =  \frac{ e^{- a_k } } { \pi } 
 \sum_{ n = 0 }^{ \infty }
\frac { a_k^n} { n!} 
 \frac{\Gamma_k}
  { ( \omega - \epsilon_k + n \, \tilde \omega_k )^2 + \Gamma_k^2  } , 
\label{eq:finalA}
\ee
where $ \epsilon_k = \Re [\varepsilon_k] $ and $ \Gamma_k = \Im [\varepsilon_k] $. 
Eq.\@ \eqref{eq:finalA} is similar to the plasmon pole version of the
CE (cf. Ref.\@ \cite{Aryasetiawan96}). However here the exponential solution
arises from a straightforward approximation to the fundamental differential
equation \eqref{eq:diffeq}: the linearization of the 
Hartree potential reveals the boson of the model (i.e., the plasmon \emph{via} peaks in 
$\mathcal{W}$), and the diagonal approximation of $\mathcal{G}$ gives rise to each isolated fermion. 
Our results are summarized in Fig.\@ \ref{fig:spf-final}. 
The dot-dashed line gives the result of this procedure together with the cross sections and the 
secondary electron background. 
The shapes of the QP peaks change little with respect to GW, but now the
full series of satellites is present. 
The internal structure of the satellites which originate from the multiple
valence bands, is also 
reproduced. This validates the decoupling approximation in the dense valence band region where, contrary to the case of an isolated core level, 
 its success is \emph{a priori} far from obvious.  
However, the intensity of the observed satellites is significantly
underestimated. 
This discrepancy is similar to that found for the CE in simple metals,
where extrinsic losses were suggested as a likely
cause \cite{Aryasetiawan96}. 
These might also be reduced by interference effects \cite{ISI:A1997WK15700033}.
To check this possibility we estimated the 
contributions from both effects 
to the satellite strengths $a_k$ using Eq.\@ 
(32) and (36) of Ref.\@ \cite{HedinMichielsInglesfield98}. 
This approach uses a plasmon pole model, Inglesfield 
fluctuation potentials, and an average over hole 
positions that takes account of the photoelectron
mean free path $\lambda$ \cite{HedinMichielsInglesfield98}. 
We observe that the averaged total satellite line shape 
in this model is similar to that for the intrinsic 
part, with a width $\gamma \approx 2$ eV due to plasmon 
dispersion. 
Thus we can approximate the extrinsic and interference 
effects by renormalizing the intrinsic satellite 
intensity, i.e., by the replacement 
${\bar a_{k}} = a_{k} + a^{ext} + a^{inf}$ in Eq.\@ \eqref{eq:finalA}. 
These quantities are evaluated with $\omega_p = 16.7$ 
eV and $\lambda=17.5$ \AA{} at 800 eV for Si, yielding 
$a^{ext}$ = 0.63 and $a^{inf} = -.11$. 
This also modifies the strength 
$Z_k=e^{-{\bar a}_k}$ of the QP peaks, but preserves 
overall normalization. 
The broadening of the satellites must also be increased, 
$\Gamma\rightarrow \Gamma + n\gamma$.
The total spectrum thus obtained (black line) is in unprecedented agreement with experiment. 
We stress that this result contains no fit parameters besides the two scaling factors 
(for spectrum and background) due to the arbitrary units of the experiment. 

The success of our present approach stresses the need to go beyond the GWA.  
The exponential representation of $\mathcal{G}$ implicitly corresponds to a vertex 
correction $\tilde\Gamma=-\frac{\delta G^{-1}}{\delta \bar \varphi}$ to the self-energy. 
Since our derivation yields $\mathcal{G}$ \emph{as a function of the screened potential } $\bar\varphi$
\eqref{eq:onelevelg}, this functional derivative can be performed explicitly, using 
$ - \frac{\delta \mathcal{G}^{-1}}{\delta \bar \varphi} = \mathcal{G}^{-1} 
\frac{\delta \mathcal{G}}{\delta \bar \varphi}\mathcal{G}^{-1} $. 
From Eq.\@ \eqref{eq:onelevelg}, a straightforward derivative of $\mathcal{G}$ contains a series of satellite contributions. 
The two inverse Green's functions lead to a significant complication, because 
they contain the inverse of this series. 
This clearly illustrates the difficulty of modelling $\tilde\Gamma$ in order to treat 
dynamical effects. It suggests rather to concentrate on modelling 
$ \frac{\delta \mathcal{G}}{\delta \bar \varphi} $, where the various
contributions are simply summed, and hence to search for a self-energy
of the form 
$ \Sigma = -i\mathcal{W}\frac{\delta \mathcal{G}}{\delta \bar \varphi}\mathcal{G}^{-1} $
instead of $\Sigma = i\mathcal{G}\mathcal{W}\tilde\Gamma$. 
In conclusion, on the basis of our experimental XPS data  we have analyzed the failure of GW to reproduce plasmon satellites and linked this 
failure to the appearance of a artificial plasmaron peak. 
On the other hand, GW results are fair when the imaginary part of $\Sigma$, hence the intensity of 
the corresponding plasmon, is small enough so that no sharp plasmaron is
created. 
Thus surprisingly, one might expect GW to work better in describing satellites
stemming from local plasmon or interband excitations close to the Fermi level
 in ``strongly correlated'' materials than for the strong plasmon
structures in conventional semiconductors. 
Starting from the fundamental equations of MBPT we have derived an exponential solution to the 
one-particle Green's function, analogous to that from the CE, that overcomes the drawbacks 
of the GWA. 
Comparison to new photoemission data shows that this yields a very good description of 
the spectral function of bulk silicon, including the satellites series. 
By calculating the secondary electron background, cross section corrections as well as a correction for extrinsic and interference effects, we achieve 
an agreement between theory and experiment that can be considered as
a benchmark. 
Our derivation also suggests how the results can be improved in cases where the
presently used approximations are inadequate. Finally,
by accessing an expression for the vertex function, our approach yields 
precious hints for directions to take in modeling dynamical effects beyond the GWA.  

\begin{acknowledgments} 
We acknowledge support by
ANR (NT09-610745), St Gobain R\&D (091986), 
and Triangle de la Physique (2007-71).
JJR and JJK are also supported in part by DOE BES Grant DE-FG03-97ER45623
and DOE CMCSN. Computer time was granted by IDRIS (544). 
We thank F.\@ Bechstedt for fruitful discussions. 
\end{acknowledgments}



%

\end{document}